\documentclass[preprint,superscriptaddress,showpacs,preprintnumbers,amsmath,amssymb]{revtex4-1}

\usepackage{amsmath}
\usepackage{amssymb}
\usepackage{graphicx}
\usepackage{morefloats}
\usepackage[usenames,dvipsnames]{xcolor}
\usepackage{blkarray}
\usepackage{verbatim}
\usepackage{hyperref}
\usepackage{color}
\usepackage{enumerate}

\begin{document}

\title{An Evolutionary Strategy based on Partial Imitation for Solving Optimization Problems}
 
\author{Marco Alberto Javarone}
\email{marcojavarone@gmail.com}
\affiliation{Dept. of Mathematics and Computer Science, University of Cagliari, Cagliari - Italy}

\date{\today}

\begin{abstract}
In this work we introduce an evolutionary strategy to solve combinatorial optimization tasks, i.e. problems characterized by a discrete search space. In particular, we focus on the Traveling Salesman Problem (TSP), i.e. a famous problem whose search space grows exponentially, increasing the number of cities, up to becoming NP-hard.
The solutions of the TSP can be codified by arrays of cities, and can be evaluated by fitness, computed according to a cost function (e.g. the length of a path). Our method is based on the evolution of an agent population by means of an imitative mechanism, we define `partial imitation'. In particular, agents receive a random solution and then, interacting among themselves, may imitate the solutions of agents with a higher fitness. Since the imitation mechanism is only partial, agents copy only one entry (randomly chosen) of another array (i.e. solution). In doing so, the population converges towards a shared solution, behaving like a spin system undergoing a cooling process, i.e. driven towards an ordered phase.
We highlight that the adopted `partial imitation' mechanism allows the population to generate solutions over time, before reaching the final equilibrium. Results of numerical simulations show that our method is able to find, in a finite time, both optimal and suboptimal solutions, depending on the size of the considered search space.
\end{abstract}

\maketitle
\section{Introduction}
In the last decades, several evolutionary algorithms~\cite{holland01,goldberg01,hofbauer01,beyer01} have been proposed to solve a wide range of optimization problems~\cite{krentel01,dorigo03,kellerer01}. Evolutionary strategies are generally based on a common scheme: a set of solutions is randomly generated then, according to specific rules, it evolves evaluating the quality of solutions by a cost/gain function.
Optimization problems may have a continuous or a discrete spectrum of solutions, defined search space. Here, we focus on discrete cases, usually referred to as combinatorial optimization problems~\cite{dorigo02}. The latter may have a huge amount of feasible solutions, identified as the minima of the cost function (or as the maxima of a gain function) of the considered problem. The size of the search space is often too large to adopt an algorithm that explores all solutions to identify the optimal one. One of the aims of quantum computing~\cite{quantum01} is to implement algorithms that are able to span, in a finite time, huge search spaces not accessible by classical algorithms. However, in the meanwhile, strategies based on the classical nature of computing play a prominent role.
In this scenario, evolutionary methods such as genetic algorithms~\cite{holland01}, and likewise other nature inspired strategies~\cite{beyer01,castro01}, allow to define heuristics that compute, in short time, a good suboptimal solution. As mentioned, in mathematical terms, an optimal solution corresponds to the global minimum/maximum of a cost/gain function. Therefore, the search space of these problems can be represented by an $n$-dimensional cost/gain function, having a continuous or a discrete domain.
In this work, we focus on the Traveling Salesman Problem (TSP hereinafter)~\cite{dorigo01}, a well-known problem with a discrete search space that, under opportune conditions, become $NP$-hard. Briefly, the TSP is based on a traveler who wants to visit a list of cities reducing the relative cost as far as possible the (e.g. the time or other resources). Hence, optimizing algorithms (e.g.~\cite{dorigo01,agliari02}) attempt to compute the best path among the listed cities. Here, the best path is the one that optimizes a function (i.e. minimizing a cost or maximizing a gain).
Increasing the number of cities, the number of feasible solutions diverges to very high values. Thus in extreme conditions even to find a good local minimum (i.e. a sub-optimum) becomes a challenging task. In particular, if every city can be visited only once, and if all cities are directly connected, the amount of feasible solutions corresponds to the factorial of the number of cities. For instance, with only $10$ cities there are $10!$ possible paths, i.e. more than $3.6 \cdot 10^6$. The solutions of the TSP can be codified by array structures containing an ordered list of cities, so that each entry of the array corresponds to a city. We aim to face the challenge of finding the optimal (or a good suboptimal) solution of the TSP by means of an agent population, whose evolution is based on an imitative mechanism described below. 
Imitation processes~\cite{brock01} have been thoroughly investigated in several scientific fields, from social psychology~\cite{aronson01} to sociophysics~\cite{galam01,loreto01} and quantitative sociology~\cite{javarone01}. For example, a wide variety of Ising-like models~\cite{mussardo01,huang01,barra01}, such as the voter model~\cite{ligget01}, allow to represent imitative mechanisms in the context of opinion dynamics~\cite{loreto01,galam02,javarone02}, evolutionary games~\cite{perc01,moreno01,tomassini01,javarone03}, and many other domains.
In the proposed model, agents of a population are provided with randomly generated solutions of a TSP. As stated above, the fitness of each solution is computed by a cost function. Then, through interacting, agents can imitate better solutions (according to fitness). Now, note importantly that the implemented imitative mechanism is partial, hence defined `partial imitation', since it entails that each agent copies only one entry of a solution better than its own. Thus, since a solution is codified by an array, `partial imitation' involves copying the value contained in just one entry.
In doing so, as we explain later, the agent population can generate solutions during its evolution and, finally, converges towards a common one (i.e. shared by all agents).
\subsection*{A statistical physics overview}
\begin{figure*}[h!]
\centering
\includegraphics[width=1.0\textwidth]{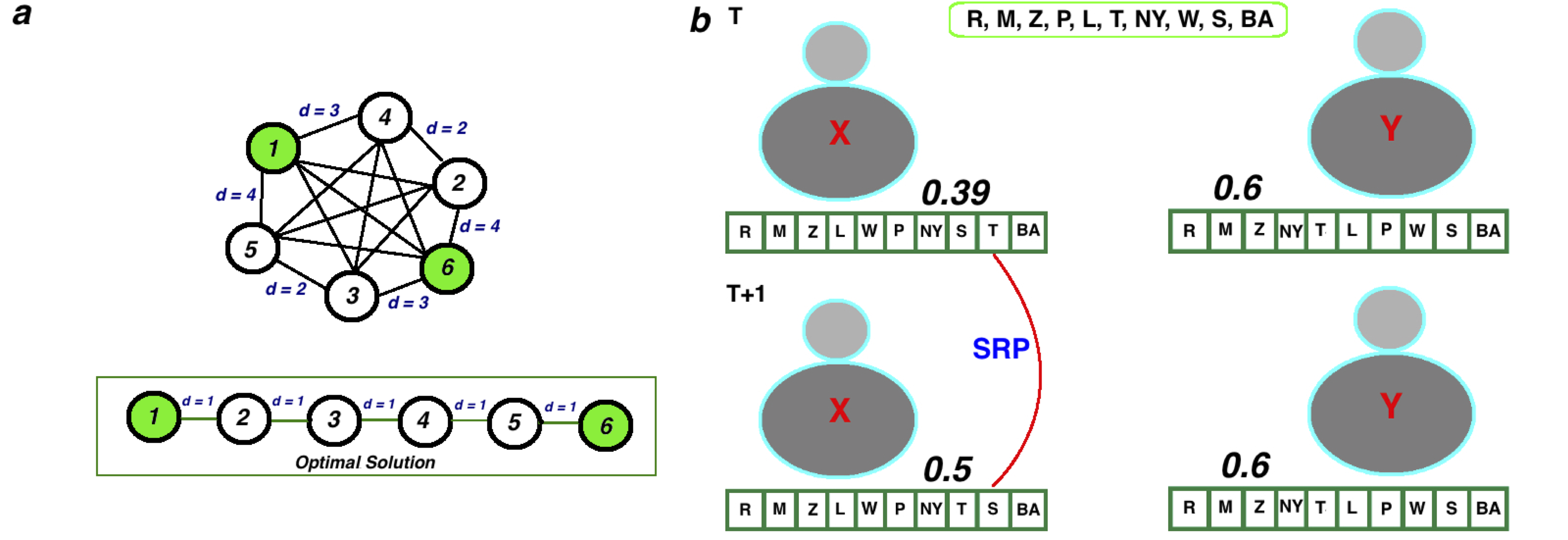}
\caption{\small \textbf{a)} A complete weighted graph~\cite{tavani01,tavani02} with six cities, with green nodes representing the origin and destination locations. The distance $d$ is computed according to the numerical value within each node. The optimal solution is shown in the lower part. \textbf{b)} Interaction between two agents, $x$ and $y$ at time $T$. The former has poorer fitness (i.e. $\eta = 0.39$), thus it undergoes a 'Solution Revision Phase'  by copying the value in the $9$th cell of the agent $y$. After this process, at $T+1$, the fitness of $x$ increased to $0.5$. Solutions are shown in the agents' arrays. The best solution is shown at the top. Each letter of the solution refers to a city, e.g. $R = Rome$, $NY = New York$, etc. The distance between the cities in the correct order is $1$. As we can see, this process allows to generate a new solution, since the updated solution of $x$ is still different from that of $y$. \label{fig:picture}}
\end{figure*}

At this point, note that our population behaves like a spin system undergoing a cooling process. This  implies that, at equilibrium, an ordered phase will be reached, with all spins aligned in the same direction. In particular, spin systems show order-disorder phase transitions~\cite{huang01} driven by temperature: at high temperatures they form a disordered paramagnetic phase, while at low temperatures (i.e. lower than a critical one, also defined as `Curie temperature') an ordered ferromagnetic phase emerges.
Order-disorder phase transitions are well studied in Ising-like models, for instance to analyze the evolution of two-opinion systems~\cite{galam03} in the presence of external influences (e.g. media), particular behaviors (e.g. conformity or stubbornness), or other attributes (e.g. gain) provided to agents.
As a result, simple two state models ($\sigma \pm 1$), can be studied analytically using the Curie-Weiss model formalism ~\cite{huang01,agliari01}, so that the phenomenon of order-disorder transitions is well described.
According to thermodynamics~\cite{huang01}, the equilibrium of a system corresponds to the minimum of its free energy $F$. In the Curie-Weiss model we have a paramagnetic phase characterized by one minimum of $F$, where both states (i.e. $\sigma = \pm 1$) coexist, and a ferromagnetic phase with two possible minima of $F$, corresponding to $\sigma = +1$ and to $\sigma = -1$. Therefore, in the ferromagnetic phase all spins are aligned in one direction. Here, the system magnetization $M$~\cite{mussardo01} is a useful order parameter that allows to directly evaluate the nature of an equilibrium.
It is defined as 
\begin{equation}\label{eq:magnetization}
M = \frac{1}{n} \sum_i s_i
\end{equation}
\noindent with $n$ number of spins.
A more complex scenario arises when studying the dynamics of spin glasses~\cite{barra02} (e.g. neural networks~\cite{barra03}). In particular, spin glasses are systems characterized by a large number of free energy minima at low temperatures. Because of the topology of spin interactions several configurations, usually referred to as patterns~\cite{amit01}, can be reached at equilibrium. In these systems, the concept of order-disorder phase transition is a little more complex. In particular, an ordered state does not correspond to the simple series of aligned spins, but to a particular pattern $\epsilon$, e.g. $\epsilon = [+1, +1, -1, +1, -1]$.
As discussed before, system magnetization makes it possible to detect the nature of a system state~\cite{mobilia01}. Thus values of $M$ close to zero indicate that the system is in a disordered (paramagnetic) phase, while values of $M = +1$ and $M = -1$ mean the system is in an ordered (ferromagnetic) phase.
Now, by using the so-called Mattis gauge, we can define a magnetization $m$ that evaluates whether a system is ordered according to a particular pattern. The Mattis magnetization reads
\begin{equation}\label{eq:mattis}
m = \frac{1}{n} \sum_i \epsilon_i s_i
\end{equation}
\noindent with $\epsilon_i$ value in the $i$-th position of the pattern, $s_i$ value of the spin in the same position of a signal $S$ of length $n$. As we can observe, when spins are perfectly aligned with a pattern $\epsilon$, the Mattis magnetization is $1$.
Note that our population cannot be construed as a proper spin glass, as agents interact, while the spins constitute the solution provided to each agent, who thus do not interact. However, the solution of a TSP can be viewed as a pattern; therefore when known, it is possible to define a kind of Mattis magnetization for the considered problem. Moreover, although spins do not interact in our model, the cost function of the TSP can be viewed in terms of an energy landscape~\cite{amit01}.
In particular, the convergence of all agents to the same solution, due to the `partial imitation' mechanism, may correspond to the convergence towards an ordered state of a spin system (as described above). From this perspective, reducing the temperature of a population, with random solutions, leads agents to form an ordered phase at equilibrium, as the latter corresponds to a pattern (i.e. a feasible solution) of the TSP. Moreover, the `partial imitation' mechanism, corresponds to a `slow cooling' process in a spin system, since by 'full imitation' the convergence towards a common solution would be faster. On the other hand, adopting a `fully imitative' mechanism would make it impossible to find the optimal solution of a problem that has not been randomly generated at the outset, due to the inability of the population to generate new sequences.
Hence, in our model, the `partial imitation' is the key to generating new solutions over time---see Figure~\ref{fig:picture}.

Now, an interesting question arises: `Is it possible to lead agents to converge towards the optimal (suboptimal) solution of a TSP?'. In our hypothesis, the slowness of the cooling process may result in agents falling into the right free energy minimum (i.e. to converge to the optimal solution). The results of our investigations show that, in the considered search spaces, our population reaches the 'optimal equilibrium'.
Before proceeding, it is important to mention other approaches to optimization problems based on statistical physics~\cite{zecchina01,krzakala01,zamponi01}. For instance, in~\cite{hopfield01} the authors analyze optimization problems by a neural approach while the work~\cite{kirkpatrick01} focuses on the `physical inspired' method of simulated annealing for solving optimization tasks. In in~\cite{mezard01} the authors study the TSP in terms of statistical mechanics; and, lastly, the authors ~\cite{bonomi01} investigate a variation of the TSP, adopting an approach based on statistical mechanics and on the metropolis algorithm.
The remainder of the paper is organized as follows: Section~\ref{sec:model} introduces the proposed model, describing in detail our heuristic based on the 'partial imitation' mechanism. Section~\ref{sec:results} shows the results of numerical simulations. Lastly, Section~\ref{sec:conclusions} ends the paper.
\section{Model}\label{sec:model}
The proposed strategy, devised for solving combinatorial optimization problems such as the TSP, is based on the simple mechanism of imitation~\cite{brock01}.
In particular, we consider a population of $N$ agents and randomly assign them a solution for a given problem. For instance, in the case of the TSP, a list of cities need to be ordered forming the shortest path, and agents know only the origin and destination cities.
Hence a solution can be represented as an array of cities, whose order corresponds to that followed by the agent during his/her path. Every solution has fitness computed according to a cost function, e.g. in the TSP the cost may correspond to total distance traveled. Thus, solutions can be compared by their fitness to identify the optimal one. In doing so, we let the population evolve following simple dynamics: at each time step, two randomly chosen agents having different solutions are selected; then, the one having the lower fitness `partially' imitates the solution of the other agent. 
As stated above, a solution is composed of several values grouped into an array. So `partial imitation' means that an agent imitates only one single value (i.e. one entry of the array) of the solution belonging to the other agent.
At first glance, this mechanism (i.e. partial imitation) seems only to slow the process of convergence towards a common solution. On the other hand, it allows to generate new solutions, as depicted in Figure~\ref{fig:picture}.
Once the fitness $\eta$ of two randomly selected agents, say $x$ and $y$, has been compared, the first one (i.e. the $x$th agent) undergoes a `Solution Revision Phase' (SRP hereinafter). It is worth stressing that the $x$th agent undergoes the SRP phase even when its fitness equals that of the $y$th agent (i.e. $\eta_x \le \eta_y$)such that the convergence towards only one final solution is not terminated when several solutions with the same fitness emerge.
The SRP may be viewed as a strategy revision phase in evolutionary games~\cite{tomassini01}. For instance, when a population evolves according to interactions based on the Prisoner's Dilemma~\cite{perc01}, agents modify their strategy by following that adopted by a richer opponent. Here, the updating strategy is performed by implementing a revision phase where the payoff used in evolutionary games corresponds to the fitness assigned to TSP solutions.
Note that, as in swarm intelligence algorithms, the number of agents in our population is constant over time; so the evolution reaches an equilibrium when all agents share the same solution.
Summarizing, the main steps of the proposed model are:
\begin{enumerate}
\item Define a population with $N$ agents, and assign each one a random solution for the considered TSP;
\item Compute the fitness of each agent (i.e. the goodness of its solution);
\item Compute the number of different solutions (say $K$) in the population;
\item IF $K > 1$:\\ \_\_$(i)$ randomly select two agents ($x$ and $y$) having different solutions:\\
\_\_\_\_ IF $\eta_x \le \eta_y$ perform the SRP (see below);\\ \_\_\_\_ ELSE REPEAT from $(i)$. \\ \_\_$(ii)$ REPEAT from $(3)$;\\\\
ELSE STOP.
\end{enumerate}
The SRP is summarized as follows
\begin{enumerate}[a]
\item Randomly select a position, say $z$, (i.e. an entry in the solution array) of the $x$-th agent's solution;
\item Check that the value in $z$ be different between the two selected agents, otherwise repeat from $(a)$;
\item Compute the position, say $w$, in the $x$-th agent's solution containing the value in position $z$ of the $y$-th agent's solution;
\item Exchange in the $x$-th agent's solution the values contained in positions $z$ and $w$.
\end{enumerate}
Consequently, in order to ensure the convergence of the algorithm towards an equilibrium, with only one shared solution, when two agents interact ---say $x$ and $y$, the former partially imitates the latter if its fitness is poorer or equal to that of $y$. 
In principle, to demonstrate that an interacting system is able to converge towards an ordered phase, an opportune mathematical demonstration is required.
For instance, in spin systems the symmetry of interactions has been shown to satisfy this relevant point. In our case, the scenario is more complex as interactions involve agents and not spins. 
Even if our interactions are symmetrical, at each time step the interaction is directed from one agent to another; thus the symmetry emerges as the possibility that, at a further step, the opposite interaction may occur (i.e. between the same two agents, but in reverse order). 
The latter observation makes our system ergodic, so that convergence is guaranteed. However, as a future work, we will focus on the identification of a formal mathematical demonstration of the convergence of our model.
\section{Results}\label{sec:results}
\begin{figure*}[h!]
\centering
\includegraphics[width=1.0\textwidth]{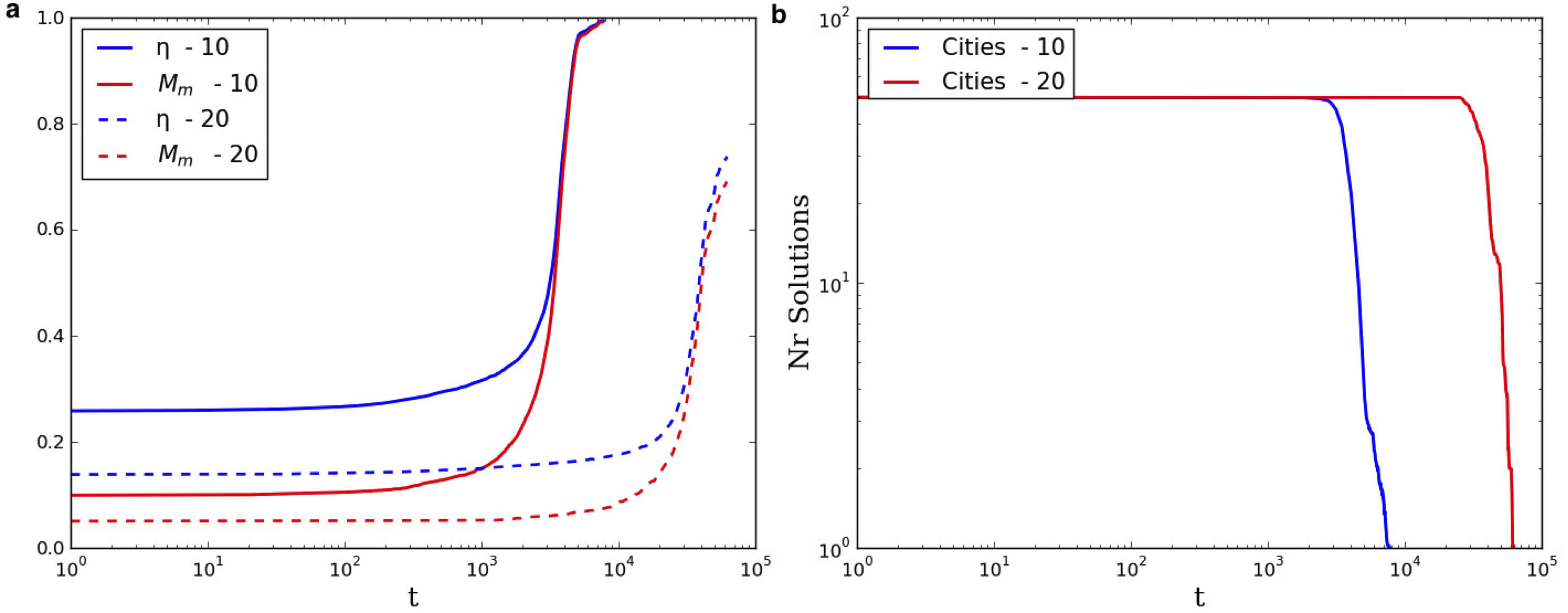}
\caption{\small \textbf{a)} Average Fitness (blue lines) and average Mattis magnetization (red lines), obtained for a population with $N = 50$ agents, in two different examples of the TSP: $10$ cities (continuous line) and $20$ cities (dotted line). \textbf{b)} Number of solutions over time for two different examples of the TSP: $10$ cities (blue line) and $20$ cities (red line), both obtained for a population with $N = 50$ agents. Results are averaged over $50$ different simulation runs.\label{fig:evolution}}
\end{figure*}

The TSP can be defined in several ways, e.g. on varying parameters such as the number of cities, introducing spatial constraints as connections based on a graph structure, and considering various features that may affect the cost function of the problem.
Here, we implement simple examples of TSPs with no spatial constraints, so that all cities are connected to each other. There is only one optimal solution, and agents know both the city of origin and destination. Although the cities are interconnected, the length of connections may vary. In particular, without loss of generality, the minimum distance is set to $1$; so the shortest path solving the problem corresponds to a series of unitary jumps.
For example, given four cities such as Rome, Paris, London and New York, whose best solution is: Rome, New York, London and Paris, the distance between Rome and New York is $d_{R,NY} = 1$, and is equal to the distance from London and Paris (i.e. $d_{L,P} =1$). As a result we consider the variant of the problem known as 'planar TSP'.
Obviously, in the provided example, distances have not been assigned in relation to their geographical values.
The setup of our TSP has as degree of freedom the number of cities $Z$, so the shortest path $D = \sum_i d_i$ is equal to $Z-1$. Therefore, knowing in advance the solution we can study the evolution of the population by considering as order parameter its Mattis magnetization (see equation~\ref{eq:mattis}). 
As mentioned above, the classical Mattis magnetization ($M_m$) refers to binary patterns, e.g. $\epsilon =[-1, +1, +1, +1, +1,-1,-1, +1]$. Thus when a signal $S$ is equal to $\epsilon$, one obtains $M_m = 1$. In the case of the TSP, we have to define an opportune function $F(\epsilon,S)$ to achieve the same result. 
Thus $F = 1$ for $\epsilon_i = s_i$ and $F = -1$ in the opposite case (i.e. $\epsilon_i \not= s_i$). Hence, we can define the Mattis magnetization of our system as
\begin{equation}\label{eq:mattis_tsp}
M_m = \frac{1}{Z} \sum_i F(\epsilon_i s_i)
\end{equation}
\noindent with $\epsilon_i$ value in the $i$-th position of the correct pattern $\epsilon$ (i.e. solution), $s_i$ value in the same position of the signal $S$ (i.e. the path) computed by an agent, and $L$ length of the pattern (i.e. number of cities).
As a result, the average value of the Mattis magnetization $<M_m>$ allows us to study the evolution of our population in relation to the exact solution ---see plot \textbf{a} of Figure~\ref{fig:evolution}. 
The proposed method has been tested varying the amount $N$ of agents in the population, in order to identify a relation between $N$ and $Z$ (i.e. the number of cities to be visited). In particular, we consider values of $Z$ between $5$ and $50$ and, since each city can be visited only once, the total number of solutions (forming the search space) corresponds to $(Z-2)!$ (recalling that agents know the city of origin and destination). 
For $Z = 50$ the search space is composed of a number of solutions of order $5 \cdot 10^{61}$, i.e. an enormous space.
Here, the fitness of each solution reads
\begin{equation}\label{eq:fitness}
\eta = \frac{Z-1}{D}
\end{equation}
\noindent and it takes values $\eta \le 1$. Lastly, a further parameter of interest is the number of different solutions over time, that at equilibrium goes to $1$ ---see plot \textbf{b} of Figure~\ref{fig:evolution}.
The plot \textbf{b} in Figure~\ref{fig:evolution} shows that an agent population spends about half its time interacting without converging to the same solution, i.e. many solutions are generated during this phase. Then, after a substantial number of interactions the total number of solutions decays to $1$. In addition, note that increasing the search space (i.e. using paths of longer length) a population, with fixed $N$, requires more time steps before starting to reduce the number of solutions. Furthermore, we see that in the final part of the evolution, the average Mattis magnetization and the average fitness almost overlap.

\begin{figure*}[h!]
\centering
\includegraphics[width=1.0\textwidth]{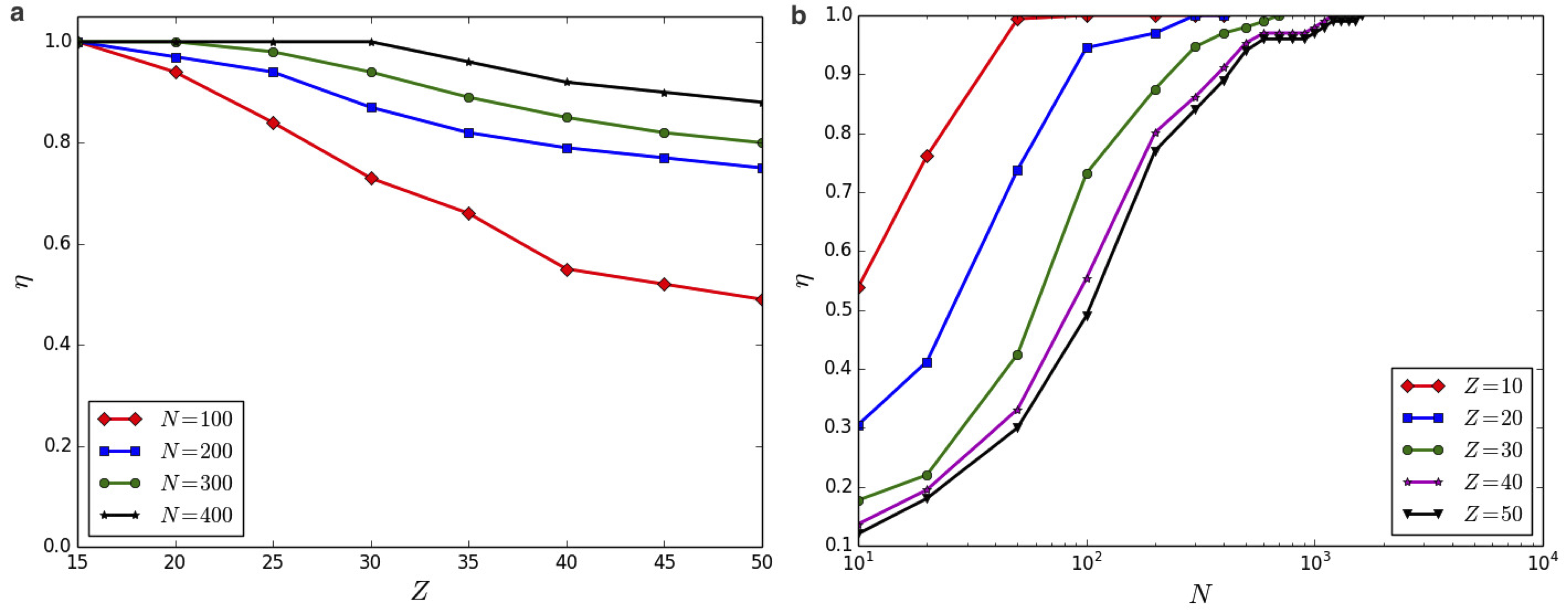}
\caption{\small \textbf{a)} Average fitness versus number of cities $Z$ for different population sizes $N$. \textbf{b)} Fitness of final solutions versus $N$ for different values of $Z$ (i.e. the number of cities). Results are averaged over $50$ simulation runs. \label{fig:fitness}}
\end{figure*}

\begin{figure*}[h!]
\centering
\includegraphics[width=1.0\textwidth]{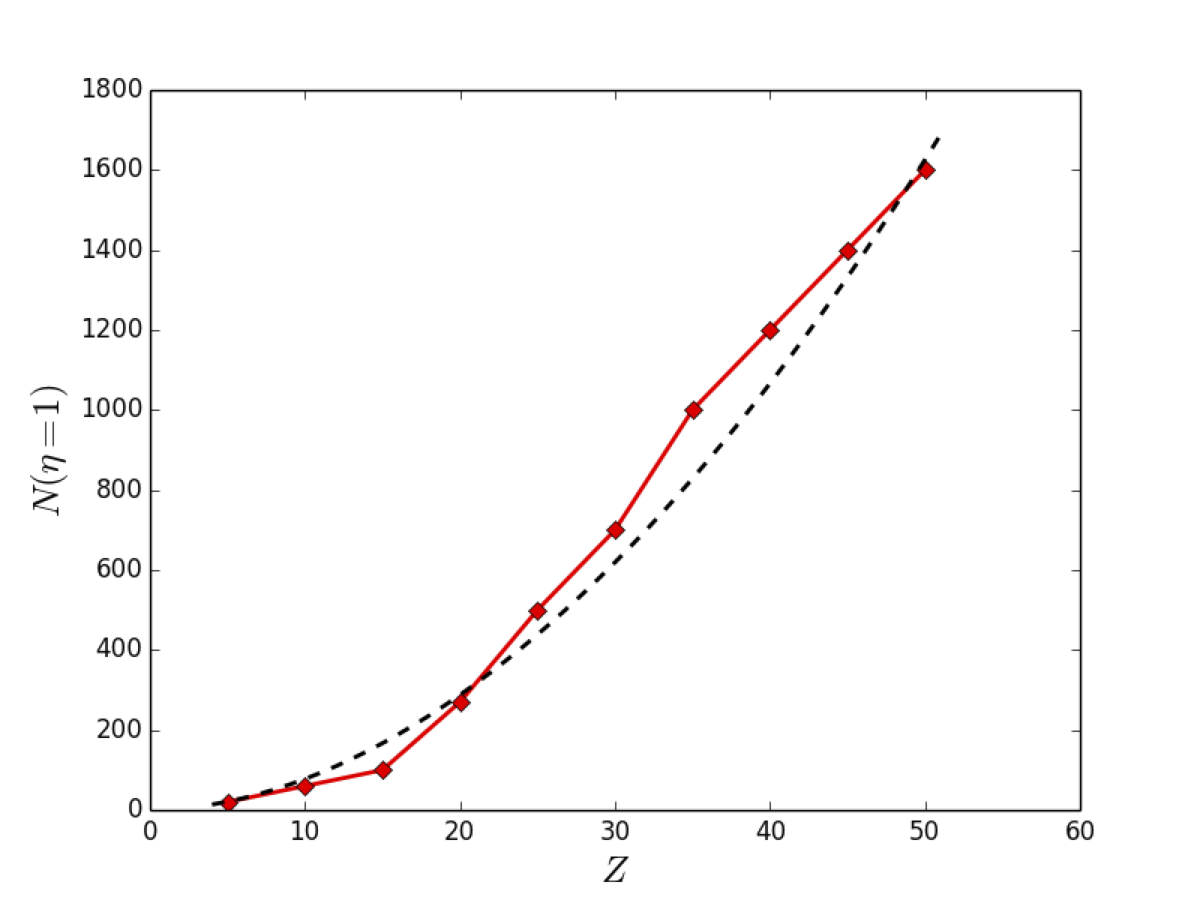}
\caption{\small Number of agents $N$ to compute the optimal solution (i.e. that with $\eta = 1$) versus path length $Z$, i.e. the total number of cities to be visited. The black dotted line fits the relation between $N_{(\eta = 1)}$ and $Z$. Results are averaged over $50$ simulation runs. \label{fig:optimum}}
\end{figure*}

Analyzing the average fitness of solutions computed varying the population size (see Figure~\ref{fig:fitness}), we clearly note that increasing $Z$ requires increasing $N$ to find the global optimum. 
Moreover, starting values of the fitness (i.e. $<\eta(0)>$) decrease as $Z$ increases, since the search space grows exponentially. It is worth noting that, for each simulation, we verified that the global optimum was not generated by the initial random assignment; hence it emerges through agent interactions (i.e. according to the `partially imitative' mechanism).
The results shown in Figure~\ref{fig:fitness}, should be interpreted as follows: on the left (i.e. plot \textbf{a}) the plot indicates that increasing $Z$ the final average fitness reduces for the same population; on the right (i.e. plot \textbf{b}), the plot shows that an opportune amount of agents is always able to find the best solution of the TSP. 
Considering only the outcomes of each run, we find that even small populations are able to solve a TSP with many cities.
For instance, the case $Z=50$ has always been exactly solved using $N = 1600$ agents, but sometimes also a population with $N = 400$ agents found the optimal solution. Furthermore, a good suboptimal solution is easily achieved also for populations with $N = 200$ agents.
Once this point has been clarified, we focus on the nature of the relation between $Z$ and $N$, in order to evaluate for each TSP the minimum number of agents required to solve the problem exactly. In Figure~\ref{fig:optimum}, the relation between the two parameters (i.e. $Z$ and $N$) is identified.
In particular, the best solution (i.e. $N_{(\eta = 1)}$) is achieved for $N$ that scales with $Z$ according to equation:  $N_{(\eta=1)}(Z) = Z^{1.89}$. Thus, by this relation we are, in principle, able to know the number of agents required to solve TSPs with many more cities.
Finally, as was to be expected, and as briefly mentioned above, the time required to reach the equilibrium phase increases with both $N$ and $Z$.
\begin{figure*}[h!]
\centering
\includegraphics[width=1.0\textwidth]{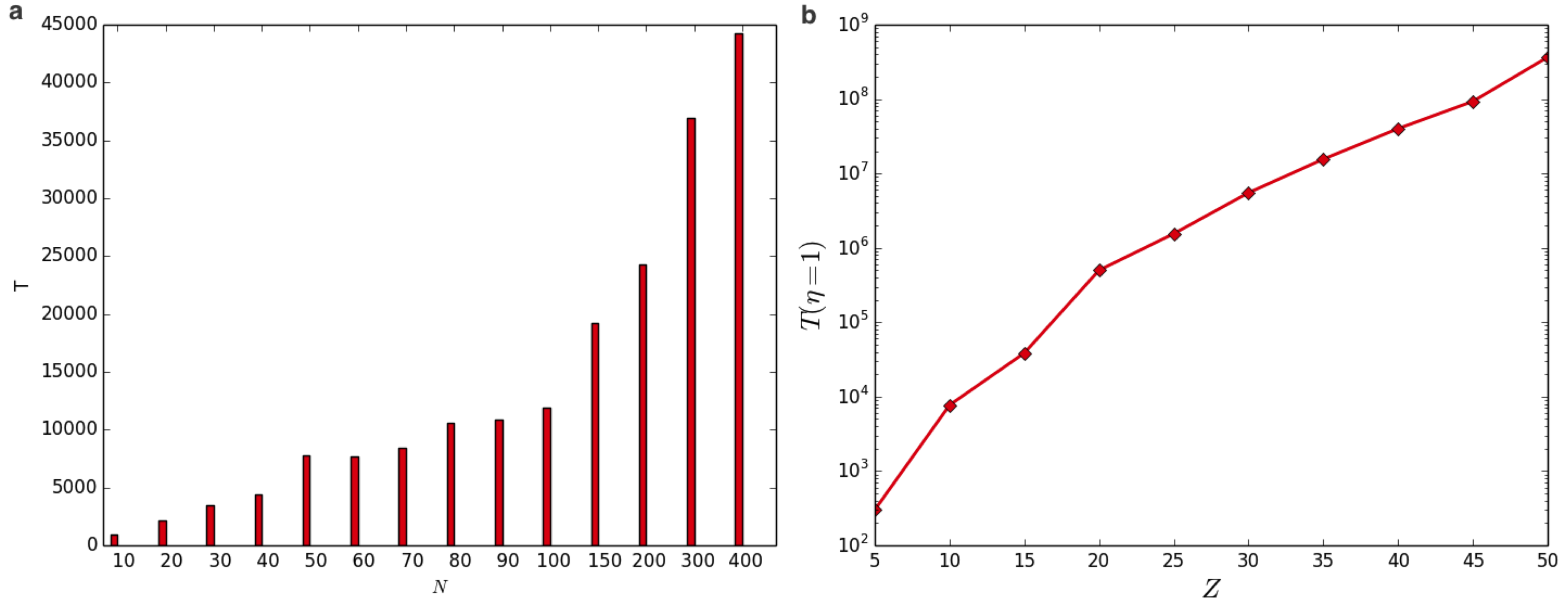}
\caption{\small \textbf{a)} Number of time steps required for the population to converge to the same solution, varying $N$ (considering the same TSP with $Z = 10$). \textbf{b)} Number of time steps required by the minimum number of agents to solve the TSP (on varying $Z$). Results are averaged over $50$ simulation runs. \label{fig:time}}
\end{figure*}
In particular, plots \textbf{a} of figure~\ref{fig:time} shows to what extent time increases, while increasing $N$ and keeping $Z$ fixed; plot \textbf{b} shows the number of time steps required to solve the TSP varying $Z$, using the minimum number of agents for every problem (e.g. $N = 100$ for $Z = 15$).
\section{Discussion}\label{sec:conclusions}
\begin{figure*}[h!]
\centering
\includegraphics[width=1.0\textwidth]{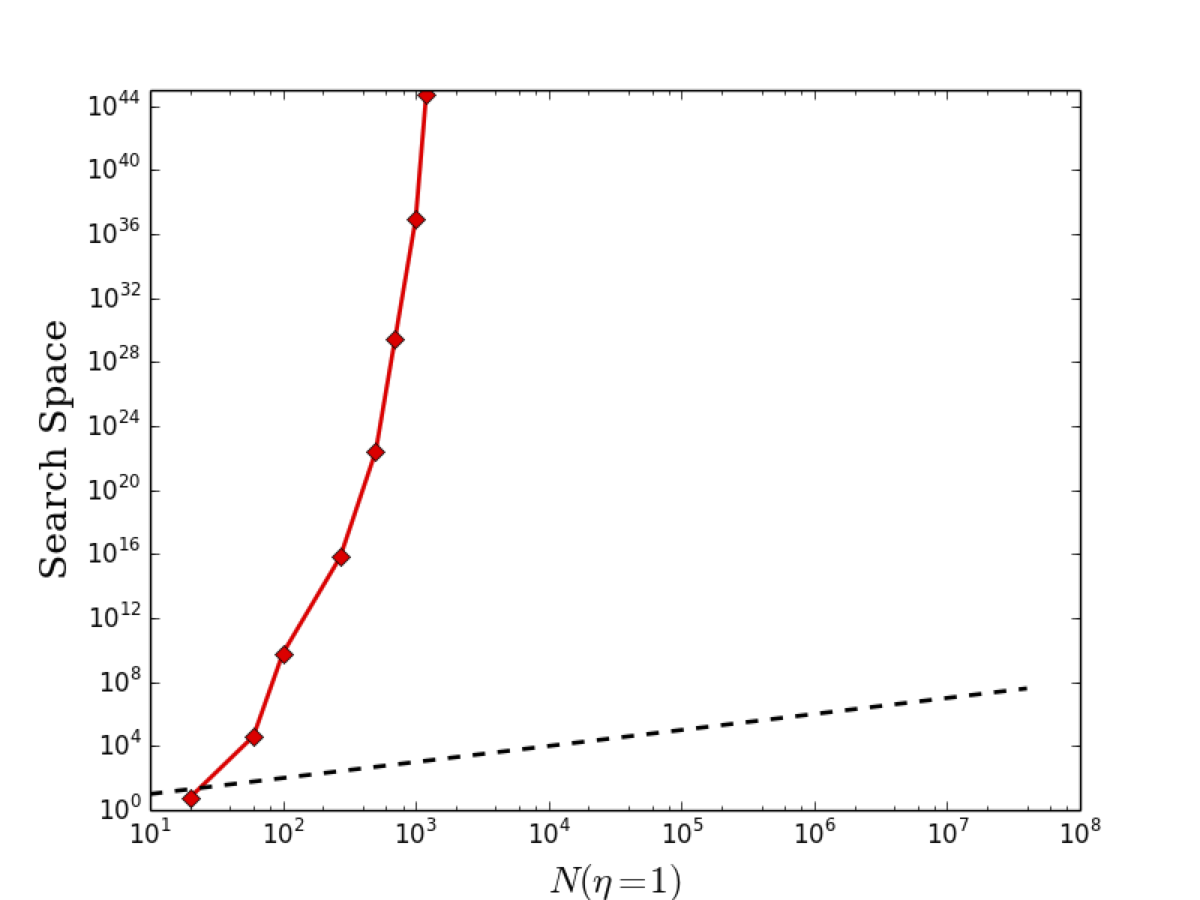}
\caption{\small Number of agents versus search space size. The black line indicates a linear relation between the search space and the number of agents. The red line indicates the relation identified in our strategy. In particular, note that the `partial imitation' mechanism  reduces substantially the number of agents as a function of the search space of a TSP. \label{fig:comparison}}
\end{figure*}

In this work, we propose an evolutionary strategy for solving combinatorial optimization problems. In particular, we focus on the famous Traveling Salesman Problem that, even in the considered case (i.e. planar), may become $NP$-hard.
Our method is based on an agent population whose interactions allow the system to evolve towards an ordered equilibrium, i.e. a state characterized by only one solution, shared among all agents.
In particular, interactions result in an imitative process between two agents: the first agent evaluates whether its fitness is equal to or smaller than that of the second one and, if this holds true, then the former copies one entry of the latter's solution. Thus, the imitative mechanism is partial and, at the same time, constitutes the way agents may generate original solutions.
As reported in figure~\ref{fig:optimum}, our population is able to exactly solve TSPs up to $50$ cities. Moreover, results show a relation between the minimum number of agents required to exactly solve a TSP and the number of cities. Furthermore, when populations are too small to always compute the optimal solution, they are able both to compute the optimal result in single attempts and to find a good suboptimum. 
Note that, in principle, when agents are provided with random solutions a TSP may be solved by using a number of agents equal to that of feasible solutions (i.e. trying to cover the whole search space). Indeed, since we are dealing with a problem that may easily become $NP$-hard, using a number of agents $N = (Z-2)!$ (with $Z$ number of cities) makes this choice unfeasible even when considering only a few cities.
Therefore, the role of interactions is to allow a population to compute an optimal (or a good suboptimal) solution with a number of agents $N \ll Z!$. 
As shown in figure~\ref{fig:comparison} interactions, based on the `partial imitative' mechanism, allow the population to solve a problem by using a number of agents much smaller than that of feasible solutions.

Importantly, note that our method may be adopted not only to solve directly an optimization problem, but also to set up a preliminary population of solutions that, later, undergo the dynamics of other heuristics (e.g. a genetic algorithm~\cite{goldberg01}). Often one of the problems of genetic algorithms is how to define an opportune initial population of solutions~\cite{initialization01,initialization02}.

Lastly, a few remarks in order to compare the underlying logic of our strategy with those of other methods adopted in optimization problems. In general, in the case of genetic algorithms~\cite{goldberg01}, interactions make it possible to perform a crossing-over mechanism and self-interactions a mutation mechanism. 
Both crossing-over and random mutations are defined in order to generate solutions with a better fitness, while reducing the amount of solutions over time. Instead, in the case of swarm intelligence based algorithms, such as ant swarms~\cite{dorigo02,dorigo03} and particle swarms~\cite{chen01}, a population explores the search space in an attempt to find, according to a set of rules, the best global (or a good local) minimum without reducing their size (i.e. the number of ants or particles). Hence, we note that our strategy is closer to that implemented by swarm intelligence algorithms.
Moreover, it is important to mention a recent optimization strategy, based on opinion dynamics processes ---see~\cite{kaur01}, that shares some similarities with our method.
Finally, we observe that beyond the `social' interpretation of our method (i.e. `partial imitation'), a further reflection warrants attention. The proposed dynamics may be thought of as that of an agent population whose interactions are based on the communication of strings (e.g. arrays) by a low quality channel. In particular, the latter only sends the fitness of a solution and one bit (i.e. one value of the considered TSP). Thus, each agent modifies its solution if the received fitness is of higher quality than its own, and if the selected bit is different from the one it uses (in the same position of the array). Therefore, although perfect communication channels are relevant in a wide range of tasks, and all problems related to the Shannon entropy~\cite{shannon01} are addressed in order to reduce noise as far as possible~\cite{tarokh01}, for generating innovative solutions, sometimes noise and disturbed signals may, counterintuitively, constitute a valuable resource. In particular, by adopting a mechanism that is completely imitative, any new solution beyond those already generated in the first step (i.e. at $t= 0$) could not arise.
Although the proposed method has been able to solve the considered TSP with a limited number of cities, we stress that it computed the optimal solution in all search spaces. Thus, our strategy may constitute the basic dynamics for implementing faster methods able to exactly solve TSPs with more cities in a finite time. Furthermore, even if we considered a simple scenario of the planar TSP, our method might be adapted to solve scenarios with direct and indirect routes. To conclude, the adopted approach shares similarities with the dynamics of spin glasses, therefore a further statistical physics analysis might, in principle, lead to gaining further insights to improve its quality.

\section*{Acknowledgments}
The author wishes to thank Adriano Barra and Elena Agliari for their helpful comments and suggestions. This work was partially supported by National Group of Mathematical Physics (GNFM-INdAM).

\end{document}